\documentclass[apjl]{emulateapj}

\newcommand{\cmjj}{\mbox{${\rm cm^{-2}}$}}
\newcommand{\hI}{\mbox{${\rm H\ I}$}}
\newcommand{\lya}{\mbox{${\rm Ly}\alpha$}}

\providecommand{\kms}{\,\ensuremath{\rm{km\,s}^{-1}}}

\providecommand{\dndz}{\ensuremath{\mbox{d${\cal N}$/d$z$}}}
\providecommand{\dndznum}{$10.4\pm 2.2$}
\providecommand{\nabs}{27}

\providecommand{\OmegaOVInum}{\ensuremath{(1.7 \pm 0.3) \times10^{-7}}}

\providecommand{\OVI}{\ensuremath{\mbox{\ion{O}{6}}}}

\providecommand{\OI}{\ensuremath{\mbox{\ion{O}{1}}}}

\providecommand{\Lya}{\ensuremath{\mbox{Ly\,}\alpha}}

\providecommand{\SII}{\ensuremath{\mbox{\ion{S}{2}}}}
\providecommand{\OI}{\ensuremath{\mbox{\ion{O}{1}}}}
\providecommand{\CII}{\ensuremath{\mbox{\ion{C}{2}}}}

\providecommand{\kms}{\,\ensuremath{\rm{km\,s}^{-1}}}

\providecommand{\mA}{\,\ensuremath{\mbox{m\AA}}}

\providecommand{\scriptOVI}{\ensuremath{\mbox{\scription{O}{6}}}}
\providecommand\scription[2]{\scriptsize#1$\;${\scriptsize\uppercase\expandafter{\romannumeral #2}}\relax}%
\providecommand{\etal}{\ensuremath{\mbox{et~al.}}}

\shorttitle{A STIS Survey for OVI Absorbers}
\shortauthors{Thom \& Chen}

\begin{document}

\slugcomment{Submitted to the Astrophysical Journal}

\title{A STIS Survey for O\,VI Absorption Systems at $0.12 < z\lesssim
0.5$ I.: The Statistical Properties of Ionized Gas$^{1}$}

\author{Christopher Thom and Hsiao-Wen Chen}
\affil{Dept.\ of Astronomy \& Astrophysics, 5640 S.\ Ellis Ave, Chicago, IL, 60637, U.S.A. \\ {\tt cthom,hchen@oddjob.uchicago.edu}}

\altaffiltext{1}{Based in part on observations with the NASA/ESA Hubble Space
Telescope, obtained at the Space Telescope Science Institute, which is operated
by the Association of Universities for Research in Astronomy, Inc., under NASA
contract NAS5--26555.}

\begin{abstract}

  We have conducted a systematic survey for intervening O\,VI absorbers in available echelle spectra
  of 16 QSOs at $z_{\rm QSO}=0.17-0.57$.  These spectra were obtained using HST/STIS with the E140M
  grating.  Our search uncovered a total of \nabs\ foreground \OVI\ absorbers with rest-frame
  absorption equivalent width $W_r(1031)\gtrsim 25\,\mA$.  Ten of these QSOs exhibit strong O\,VI
  absorbers in their vicinity.  Our O\,VI survey does not require the known presence of \lya, and
  the echelle resolution allows us to identify the O\,VI absorption doublet based on their common
  line centroid and known flux ratio.  We estimate the total redshift survey path, $\Delta\,z$,
  using a series of Monte-Carlo simulations, and find that $\Delta\,z=1.66, 2.18$, and 2.42 for
  absorbers of strength $W_r=30, 50$ and $80$ m\AA, respectively, leading to a number density of
  $d\,{\cal N}(W\ge 50\,\mA)/dz = 6.7\pm 1.7 $ and $d\,{\cal N}(W\ge 30\,\mA)/dz = 10.4 \pm 2.2$.
  In contrast, we also measure $d\,{\cal N}/dz=27\pm 9$ for O\,VI absorbers of $W_r>50\ \mA$ at
  $|\Delta\,v|< 5000$ \kms\ from the background QSOs.  Using the random sample of O\,VI absorbers
  with well characterized survey completeness, we estimate a mean cosmological mass density of the
  O\,VI gas $\Omega({\rm O}^{5+})\,h = \OmegaOVInum$.  In addition, we show that $<5$\% of O\,VI
  absorbers originate in underdense regions that do not show a significant trace of H\,I.
  Furthermore, we show that the neutral gas column $N(\hI)$ associated with these O\,VI absorbers
  spans nearly five orders of magnitude, and shows moderate correlation with $N({\rm O\,VI})$.
  Finally, while the number density of O\,VI absorbers varies substantially from one sightline to
  another, it also appears to be inversely correlated with the number density of \hI\ absorbers
  along individual lines of sight.

\end{abstract}


\section{Introduction}

Uncovering the missing baryons in the present-day universe is one of the central issues in
observational cosmology, because an accurate accounting of the baryon budget at different locations
sets an important constraint in modeling galaxy formation and evolution.  Over the past few years,
measurements of the baryonic mass density from independent methods have approached a consistent
value with high precision, $\Omega_b\,h^2 = 0.020 \pm 0.002$ \citep{burles-etal-01, spergel-etal-03,
  wang-etal-03, omeara-etal-06}.  But while $> 95$\% of all baryons can be accounted for in the
\lya\ forest at redshift $z\sim 3$ \citep{rauch-etal-97}, only $1/3$ have been identified at
redshift $z=0$ in known components, such as stars and neutral gas in galaxies or hot plasma in
galaxy clusters \citep[e.g.][]{persic-salucci-92, fukugita-etal-98}.

Cosmological simulations indicate that $\approx 30-40$\% of all baryons at the present epoch reside
in a diffuse warm-hot intergalactic medium (WHIM), with temperatures in the range $10^5 < T < 10^7$ K,
rather than in virialized halos where galaxies or galaxy groups lie
\citep{cen-ostriker-99-missing-baryons,dave-etal-01-WHIM-baryons}.  The simulations also show that
this WHIM is shock-heated as tenuous gas accretes onto large-scale filamentary structures and
remains hot owing to the low gas density and inefficient cooling.  Locating the WHIM in order to
verify these models is difficult, because the temperature of the gas is too low to be easily
detected with the current generation of x-ray satellites and is too hot to be observed from the
ground using optical-IR facilities.  Simple ionization modeling of hot gas at $T\sim 3\times 10^5$ K
and metallicity 0.1 solar has shown that the gaseous clouds would imprint prominent absorption
features produced by ionic transitions such as O\,VI $\lambda\lambda$ 1031, 1037 and Ne\,VIII
$\lambda\lambda$ 770, 780 in the spectra of background QSOs \citep[e.g.][]{verner-etal-94,
  mulchaey-etal-96}.  Surveys for the WHIM based on the presence of these ions at $z<1$ therefore
require a high-resolution UV spectrograph in space.

\citet{tripp-etal-00-H1821+643} searched for intervening O\,VI absorbers along the sightline toward
H1821$+$643 ($z_{\rm QSO}=0.297$) in echelle spectra ($R\equiv\lambda/\Delta\lambda\approx 43,000$)
obtained using the Space Telescope Imaging Spectrograph (STIS) on board the Hubble Space Telescope
(HST).  The search indicated that intervening O\,VI absorption systems found in QSO spectra may
indeed contain a significant fraction of baryons that had been missed by surveys based on the direct
detection of photons emitted by known objects.  Subsequent searches along individual sightlines have
been carried out toward PG0953$+$415 at $z_{\rm QSO}=0.239$ \citep{savage-etal-02-PG0953+415},
PG1259$+$593 at $z_{\rm QSO}=0.478$ \citep{richter-etal-04-PG1259+593}; PG1116$+$215 at $z_{\rm
  QSO}=0.1765$ \citep{sembach-etal-04-PG1116+215}, PKS0405$-$123 at $z_{\rm QSO}=0.573$
\citep{prochaska-etal-04-PKS0405-12}, and HE0226$-$4110 at $z_{\rm QSO}=0.495$
\citep{lehner-etal-06-HE0226-4110}.  The measured line density of intervening O\,VI absorbers with
rest-frame absorption equivalent width of the 1031 transition $W_r \ge 50$ m\AA\ ranges from $d{\cal
  N}({\rm O\,VI})/dz=11$ to 36\footnote{\citet{tripp-etal-00-H1821+643}
  reported $d{\cal N}({\rm O\,VI})/dz\sim 48$ based on four O\,VI absorbers found with $W_r > 30$
  m\AA\ toward H1821$+$643.  Three of the four absorbers have $W_r > 50$ m\AA.  We therefore scale
  $d{\cal N}({\rm O\,VI})/dz$ accordingly and work out $d{\cal N}({\rm O\,VI})/dz=36$ for O\,VI
  absorbers stronger than $W_r=50$ m\AA.} per line of sight at $z_{\rm O\,VI}=[0,0.5]$.

The large scatter in the observed number density of O\,VI absorbers suggests a significant variation
in the spatial distribution of O$^{5+}$ ions across different sightlines.  It also underscores a
potential systematic bias in evaluating the cosmological mass density of warm-hot gas based on
selective quasar sightlines.  The first blind search of O\,VI absorbers was carried out at $z_{\rm
  O\,VI}=[0.5,2.1]$ toward 11 background quasars by \citet{burles-tytler-96-OVI-Omega_b} using
medium resolution ($R\approx 1300$) and $S/N\approx 20$ Faint Object Spectrograph spectra.  They
reported $d{\cal N}({\rm O\,VI})/dz=1.0\pm 0.6$ per line of sight at a mean redshift $\langle z_{\rm
  O\,VI}\rangle=0.9$ with $W_r \ge 210$ m\AA.  More recently,
\citet{danforth-etal-05-OVI-baryon-census} conducted a search of O\,VI absorption features in {\it
  known} \lya\ absorbers at $z_{\rm abs}<0.15$, using Far Ultraviolet Spectroscopic Explorer (FUSE)
spectra ($R\approx 15,000$) of 31 active galactic nuclei (AGN).  Their search yielded $d{\cal
  N}({\rm O\,VI})/dz=9\pm 2$ and $17\pm 3$ per line of sight with $W_r \ge 50$ and $W_r\ge 30$ m\AA,
respectively, at $z_{\rm O\,VI} < 0.15$, which is lower than measurements from individual lines of
sight discussed earlier.

Despite the large scatter in the number density of intervening O\,VI absorbers, these measurements
imply that at least $\sim 10\%$ of the total baryons reside in the gaseous clouds probed by the
O\,VI transition.  Assuming a mean metallicity of 1/10 solar and an ionization fraction of 20\% for
the O\,VI absorbers, Tripp \etal\ (2000) derived a lower limit to the cosmological mass density in
the ionized gas of $\Omega_b({\rm O\,VI})\gtrsim 0.003\,h^{-1}$ and Danforth \& Shull (2005)
derived $\Omega_b({\rm O\,VI})\approx 0.002\,h^{-1}$.  This general inference of O\,VI absorbers
being a substantial reservoir of missing baryons is supported by numerical simulations that result
in a mean metallicity of $1/10$ solar for the WHIM \citep{cen-etal-01, fang-bryan-01-H1821+643_IGM}.
However, additional uncertainties arise due to the unknown ionization state and metallicity of the
gas.

We have conducted a systematic survey of intervening O\,VI absorbers in available echelle spectra
obtained using HST/STIS with the E140M grating.  A search of the HST data archive yielded
high-resolution ($R\approx 45,000$) spectra of 16 distant quasars at $z_{\rm QSO}=0.17-0.57$ with a
mean $S/N > 7$ per resolution element.  These quasars have been selected by different authors for
different studies and have bright UV fluxes that are suitable for STIS echelle observations.
Therefore, they represent a random, flux-limited sample of quasar spectra for an unbiased search of
intervening absorbers.  The entire STIS echelle spectroscopic sample of quasars summarized in Table
1 defines the random sightlines along which our blind search of intervening O\,VI absorbers was
conducted.

The primary objectives of the blind O\,VI survey are: (1) to obtain a statistically representative
sample of O\,VI absorbers along random lines of sight for an accurate measurement of the
cosmological mass density of warm-hot gas at low redshift; and (2) to study the incidence of O\,VI
absorbers along individual sightlines for quantifying the spatial variation of O\,VI absorbing gas.
These O\,VI absorbers identified along random sightlines also form a uniform sample for studying the
physical nature of tenuous gas probed by the presence of O$^{5+}$ and for investigating their
large-scale galactic environment.  Here we focus our discussion on the statistical properties of
O\,VI absorbers.  Studies of the gas properties and discussions of individual absorbers are given in
an accompanying paper (Thom \& Chen, 2008; hereafter paper~II), and studies of the surrounding
galaxies will be presented in future papers.

This paper is organized as follows.  In Section 2, we summarize the available quasar spectra and the
data quality.  In Section 3, we describe our approach for identifying an O\,VI $\lambda\lambda$
1031, 1037 absorption doublet with no prior knowledge of the presence of H\,I or other ionic
transitions.  A Monte-Carlo analysis was performed for evaluating the completeness of the survey.
In Section 4, we present a statistical sample of O\,VI absorbers identified along the sightlines
toward the 16 background quasars.  We measure different statistical quantities of the ionized gas.
In Section 5, we compare our search results with others and discuss the implications in Section 6.
We adopt a $\Lambda$ cosmology, $\Omega_{\rm M}=0.3$ and $\Omega_\Lambda = 0.7$, with a
dimensionless Hubble constant $h = H_0/(100 \ {\rm km} \ {\rm s}^{-1}\ {\rm Mpc}^{-1})$ throughout
the paper.

\section{The STIS Echelle Spectra}
\label{sec: observations}

\begin{deluxetable*}{p{1.5in}rrrcr}
\tablecaption{Summary of the STIS Echelle Spectra}
\tablewidth{0pt}
\tablehead{\colhead{QSO} & \colhead{$z_{\rm QSO}$} & \colhead{$z_{\rm min}$} & 
\colhead{$z_{\rm max}$\tablenotemark{a}} & \colhead{$t_{\rm exp}$} & \colhead{PID}  \\
\colhead{(1)} & \colhead{(2)} & \colhead{(3)} & \colhead{(4)} & \colhead{(5)} & \colhead{(6)}}
\startdata
HE\,0226-4110 \dotfill  &  0.4950 &    0.1154 &    0.4707 &   43772 &   9184  \\
PKS\,0312-77  \dotfill  &  0.2230 &    0.1241 &    0.1999 &   37908 &   8651  \\
PKS\,0405-12  \dotfill  &  0.5726 &    0.1241 &    0.5478 &   27208 &   7576  \\
HS\,0624+6907 \dotfill  &  0.3700 &    0.1222 &    0.3464 &   61950 &   9184  \\
PG\,0953+415  \dotfill  &  0.2390 &    0.1144 &    0.2163 &   24478 &   7747  \\
3C\,249.1     \dotfill  &  0.3115 &    0.1222 &    0.2885 &   68776 &   9184  \\
PG\,1116+215  \dotfill  &  0.1765 &    0.1144 &    0.1536 &   39836 &   8165/8097  \\
PG\,1216+069  \dotfill  &  0.3313 &    0.1309 &    0.3078 &   69804 &   9184  \\
3C\,273       \dotfill  &  0.1580 &    0.1144 &    0.1417 &   18671 &   8017  \\
PG\,1259+593  \dotfill  &  0.4778 &    0.1144 &    0.4533 &   95760 &   8695  \\
PKS\,1302-102 \dotfill  &  0.2784 &    0.1183 &    0.2558 &   22119 &   8306  \\
PG\,1444+407  \dotfill  &  0.2673 &    0.1222 &    0.2442 &   48624 &   9184  \\
3C\,351.0     \dotfill  &  0.3719 &    0.1309 &    0.3483 &   73198 &   8015  \\
PHL\,1811     \dotfill  &  0.1917 &    0.1144 &    0.1690 &   33919 &   9418  \\
H\,1821+643   \dotfill  &  0.2970 &    0.1144 &    0.2741 &   50932 &   8165  \\
Ton\,28       \dotfill  &  0.3297 &    0.1231 &    0.3059 &   48401 &   9184  \\
\enddata
\tablenotetext{a}{The maximum redshift is defined for O\,VI absorbers
at velocity separation $> 5000$ \kms\ from the background QSO, but the
line search is conducted through the emission redshift of the QSO.}
\end{deluxetable*}

The goal of our project is to conduct a blind survey of O\,VI $\lambda\lambda$ 1031, 1037 absorption
doublet features along random lines of sight toward background quasars for establishing a
statistically representative sample of the O\,VI absorbers at $z<0.5$.  The survey therefore
requires high-quality UV spectra of distant quasars.  In addition, we focused our search on
available echelle quality spectra in order to uncover the majority of O\,VI absorbing gas.  Both
numerical simulations presented by \citet{chen-etal-03} and the recent FUSE survey by
\citet{danforth-etal-05-OVI-baryon-census} show that the bulk ($>60$\%) of O\,VI absorbers have
rest-frame absorption equivalent width of the 1031 transition $W_r=30-100$ m\AA.  An echelle
spectrum with resolution element $\delta\,v\approx 7$ \kms\ at moderate $S/N$ ($\approx 5$) per
resolution element is therefore necessary for detecting these weak lines.  This criterion excludes
Faint Object Spectrograph data (typically $\delta\,v\approx 230-270$ \kms) from our search.
Finally, the throughput of STIS declines steeply at $\lambda < 1180$ \AA, corresponding to a minimum
$z_{\rm O\,VI}\approx 0.14$, and we therefore include in our sample only those sightlines toward
quasars at $z_{\rm QSO}\gtrsim 0.16$ for the search of intervening O\,VI absorbers.

Our search in the HST data archive yielded 16 quasars observed with the STIS E140M grating.  A
summary of the STIS observations is presented in Table 1, where we list in columns (1)--(6), the QSO
field, $z_{\rm QSO}$, minimum and maximum redshifts over which the sample of random O\,VI absorbers
is established ($z_{\rm min}$, $z_{\rm max}$), total exposure time, and the program ID.  We note
that $z_{\rm max}$ is defined for O\,VI absorbers at velocity separation $> 5000$ \kms\ from the
background QSO, but the line search is conducted through the emission redshift of the QSO.
Individual flux-calibrated spectra were processed using the latest calibration files
\citep{aloisi-etal-05} and retrieved from the MAST server.  For each QSO, individual echelle orders
were continuum normalized and co-added to form a single, stacked spectrum using our own software.
The continuum was determined using a low-order polynomial fit to spectral regions that are free of
strong absorption features.  The combined echelle spectra of 16 QSOs have on average $S/N>7$ per
pixel over the majority of the spectral regions covered by the data.  Together the total redshift
survey path length covered by the echelle sample is $\Delta\,z\approx 2.73$ over $z=[0.12,0.5]$,
comparable to the total redshift path length at $z<0.15$, $\Delta\,z=2.2$, carried out by
\citet{danforth-etal-05-OVI-baryon-census} toward 31 sightlines in FUSE spectra.  Given the varying
$S/N$ versus wavelength along and between individual sightlines, we determine the $\Delta\,z$ versus
the O\,VI 1031 absorption equivalent width threshold $W_0$ in \S\ 3.2.

\section{A Statistical Sample of O\,VI Absorption Systems}

\subsection{The Search for O\,VI Doublet Features}

To establish a statistically representative sample of O\,VI absorbers, we carry out a blind survey
of O\,VI doublet features along each sightline that does not require known presence of hydrogen
\lya\ or other ionic transitions.  Before conducting the doublet search in each QSO spectrum, we
first exclude spectral regions of known absorption features from the Milky Way.  Next, we search in
the literature for known intervening absorbers, and further exclude spectral regions that contain
these known features from the subsequent doublet search. For those sightlines not well studied, we
performed this search ourselves.  For the doublet search, we first smooth each stacked echelle
spectrum using the Hanning function.  Next, we step through the smoothed spectrum and search for
deviant pixels that show negative fluctuation from unity by $> 1.5\ \sigma_c$, where $\sigma_c$ is
the 1-$\sigma$ error per smoothed pixel.  We then determine whether a deviant pixel is a candidate
absorption line based on a single gaussian-profile fit to the negative fluctuation.  We restrict the
width of the model profile to be no less than the spectral resolution element and no larger than 300
\kms.  The last criterion is justified based on the line width distribution of known O\,VI absorbers
\citep[]{heckman-etal-02, danforth-etal-05-OVI-baryon-census}.  The deviant pixel is discarded from
further consideration if the absorption equivalent width $W$ based on the best-fit gaussian profile
has $< 2\ \sigma$ significance.

Next, we treat each candidate absorption feature as the O\,VI 1031 line and determine the
corresponding redshift.  Then we examine whether the spectrum is consistent with the presence of
both members of the O\,VI doublet based on a simultaneous doublet profile fit at the expected
locations of the redshifted O\,VI 1031 and O\,VI 1037 lines.  At this stage, we restrict both
members to share the same line width and have a fiducial line ratio of $2:1$, given the ratio of
their respective oscillator strengths.  The product of the doublet fitting procedure is a list of
candidate O\,VI absorption doublets and their rest-frame absorption equivalent widths, $W_r(1031)$
and $W_r(1037)$.  We visually examine every candidate and accept doublets that satisfy both of the
following two criteria.  First, the O\,VI 1031 line has $> 3\,\sigma$ significance.  Second, the
line ratio $R_{\rm O\,VI}\equiv W_r(1031)/W_r(1037)$ falls between $1-\sigma_{R_{\rm O\,VI}}$ and
$2+ 2\,\sigma_{R_{\rm O\,VI}}$, where $\sigma_{R_{\rm O\,VI}}$ is the associated measurement
uncertainty in $R_{\rm O\,VI}$.  Because the O\,VI 1031 transition is required to have at least
3-$\sigma$ significance, the presence of both members together warrants a higher significance for
the detection of each O\,VI doublet.

The entire sample of uncovered O\,VI absorbers include both foreground absorbers and absorbers that
are associated with the host of the background QSOs.  In the following discussion, we define a
``random sample'' that consists of absorbers with $|\Delta\,v|> 5000$ \kms\ from the background QSOs
and an ``associated sample'' that are within $|\Delta\,v|< 5000$ \kms\ of the QSOs, where the
ionization background is expected to be significantly different from the meta-galactic radiation
field \citep[e.g.][]{weymann-etal-81}.  The results of our search are summarized in Table 2.  For
each QSO line of sight, we report in columns (2) through (4) the total number of random absorbers,
the minimum absorption equivalent width of the uncovered absorbers on that sightline, $W_r^{\rm
  min}$, and the presence/absence of absorbers in the vicinity of the background QSO.

\begin{figure*}
\epsscale{1.0}
\plotone{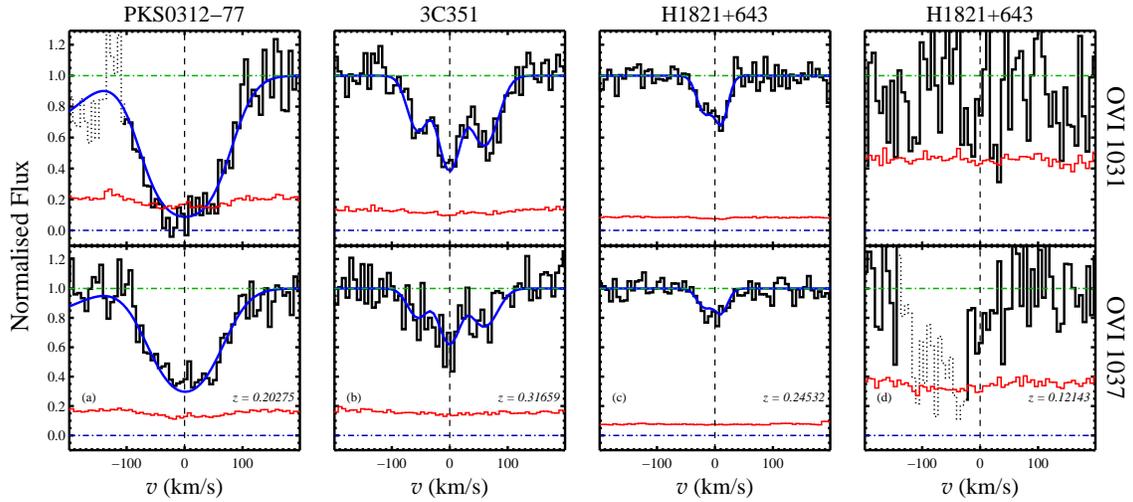}
\caption{Examples of \scriptOVI\ doublets.  For each doublet, the
  $\lambda 1031$ member is plotted in the top panel and the associated
  $\lambda 1037$ line at the bottom.  The corresponding 1-$\sigma$
  error array is shown in orange.  We also present a best-fit Voigt
  profile of each absorber in blue.  The spectra have been binned by a
  factor of 2 for display only.  In panels (a) and (b), we present new
  O\,VI absorbers uncovered at $z=0.2028$ and $z=0.3166$ in our search
  routine, along the sightlines toward PKS0312$-$77 and 3C351,
  respectively.  In panel (c), we present a previously known absorber
  at $z=0.24532$ toward H\,1821$+$643 \citep{tripp-etal-00-H1821+643}
  that has also been recovered in our blind search.  Finally, we show
  in panel (d) an absorber previously reported by these same authors
  at $z=0.1214$ toward H\,1821$+$643 based on additional FUSE spectra
  of the QSO.  It is clear that the quality of available STIS spectra
  does not allow us to recover this absorber.  Our random sample
  therefore does not include this system. }
\label{fig: absorber_plots}

\end{figure*}

Figure 1 presents three examples of the O\,VI absorption doublet uncovered in our blind search
(panels a--c), together with an example of a previously reported absorber that is not recovered in our
search (panel d).  While our search routine has uncovered \nabs\ absorbers toward 16 random lines of
sight, four of 16 systems previously reported toward seven lines of sight by other authors are not
confirmed in our survey.  We discuss these systems in more detail in \S\ 5 and paper~II.  As a
final step, we identify additional transitions associated with each fiducial O\,VI absorber.  We
perform a Voigt profile analysis to the \OVI\ doublets, as well as $\mbox{Ly}\,\alpha, \beta$ (where
available) using the VPFIT\footnote{http://www.ast.cam.ac.uk/~rfc/vpfit.html} software package. In
cases where absorbers have a complex structure, we attempt a multiple-component model profile.  Note
that we consider absorption components with velocity separation $\Delta\,v \le 200$ \kms\ as a
single absorber in the statistical sample.

\subsection{Completeness of the Search}

\begin{figure*}
  \epsscale{1.0} 
  \plotone{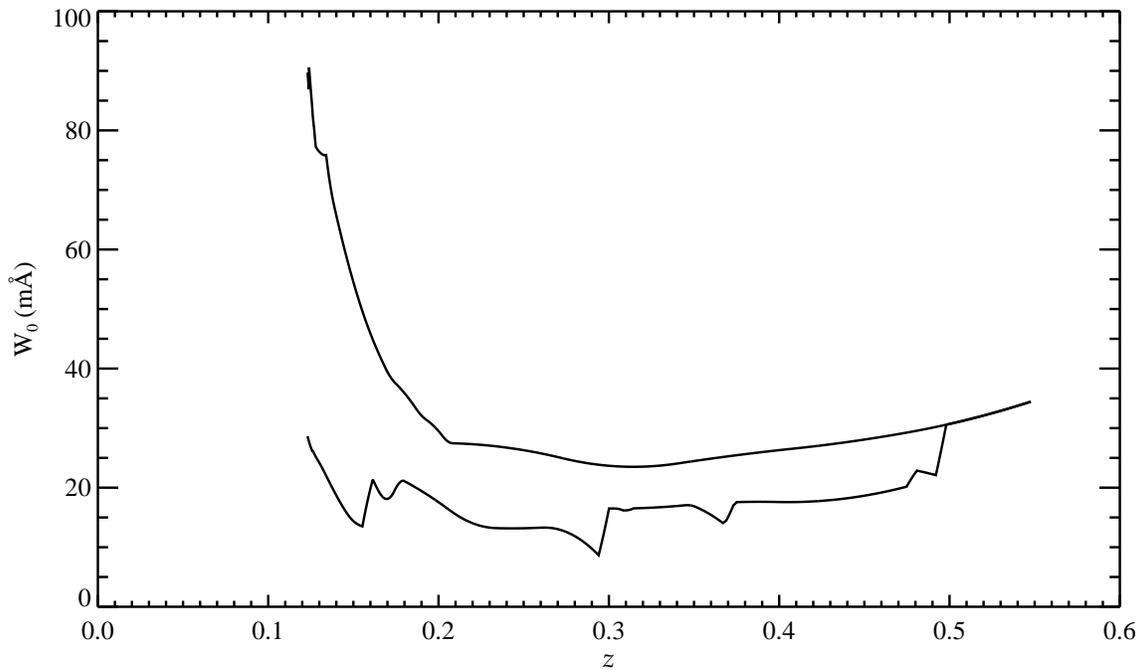}
  \caption{The 99.7\% completeness limit in the absorption equivalent
    width, $W_0$, of the O\,VI 1031 transition versus redshift, for
    all 16 sightlines in the STIS echelle sample.  At each redshift,
    the spread in $W_0$ represents varying spectral quality across
    different lines of sight.  For absorbers at $z_i$ with
    $W_r>W_0(z_i)$, they are expected to be detected at greater than
    3-$\sigma$ confidence level.  The plot shows that the STIS echelle
    sample as a whole is most sensitive to $W_r \sim 30$ m\AA\
    absorbers at $z\approx 0.2-0.5$. while at $z<0.2$ the majority of
    the spectra are sensitive to absorbers of $W_r>50$ m\AA.}
  \label{fig: ew}
\end{figure*}

To characterize the statistical properties of the O\,VI absorbers uncovered in our blind doublet
search, it is necessary to first estimate the completeness of the survey.  This is particularly
important for the STIS data because (1) the $S/N$ varies substantially across the entire wavelength
range in each echelle spectrum and (2) individual QSO spectra were obtained from different programs
and span a wide range in their mean noise levels.  We perform a suite of Monte-Carlo simulations to
quantify the completeness of our blind doublet search.  A specific goal of the Monte-Carlo
simulations is to estimate the total redshift path length $\Delta\,z$ covered by the survey for a
given absorption equivalent width threshold in the O\,VI 1031 transition, $W_*$, over which
absorbers of $W_r\ge W_*$ can be recovered.  Exploring the completeness versus $W_*$ allows us to
maximize the survey efficiency, given the already small spectroscopic quasar sample.

To carry out the simulations, we first mask regions along a single sightline $k$ that are blocked by
strong Galactic or extra-galactic absorption (e.g.\ Galactic O\,I\,1302, or Galactic and
extra-galactic \Lya).  Note that a strong absorption line at $\lambda_x$ will render us insensitive
to doublet features in {\it two} regions in redshift space ($\lambda_x/1031.927 - 1.0$ and
$\lambda_x/1037.616 - 1.0$), since it may mask either member of the \OVI\ doublet.  Next, we step
through the redshift space defined by $z_{\rm min}$ and $z_{\rm QSO}$ of the sightline defined in
Table 1.  At each redshift $z_i$, we adopt an intrinsic line width of $b=10$ \kms\ (corresponding to
a gas temperature of $T\sim 10^5$ K), generate a model absorption line of varying strength $W_{in}$,
redshift the model to $z_i$, and perturb the smooth model within the random noise characterized by
the 1-$\sigma$ error array of the original echelle spectrum.  Next, we apply the line search program
and evaluate whether the synthetic absorption feature can be recovered.  We repeat the procedure
1,000 times for each input line strength and determine the 3-$\sigma$ detection threshold at
$W_0^k(z_i)$, where the recovery rate of the synthetic absorption features is 99.7\%.  Finally, we
repeat the procedure for all 16 QSO spectra in the echelle sample.

The result of the Monte-Carlo simulations is a sensitivity curve $W_0^k(z)$ for every QSO sightline
$k$ that represents the 99.7\% completeness of our blind O\,VI survey.  These curves together allow
us to calculate the total survey path length versus $W_0$ for the entire sample of O\,VI absorbers
from our search.  For absorbers of $W_r>W_0$, they are expected to be detected at greater than
3-$\sigma$ confidence level.  We present in Figure 2 the collective $W_0(z)$ curve from all 16 QSO
sightlines together.  At each redshift, the spread in $W_0$ represents varying spectral quality in
the echelle spectra of different QSOs.  The plot shows that our blind search is sensitive to $W_r
\sim 30$ m\AA\ absorbers at $z\approx 0.2-0.5$, while at $z<0.2$ the majority of the spectra are
sensitive to absorbers of $W_r>50$ m\AA.

To calculate the total survey path for our blind search, we first define a redshift path density
$g(W_r,z)$ at $z_i$ as
\begin{equation}
g(W_r,z_i)=\sum_k\,H(z_i-z_{\rm min}^k)\,H(z_{\rm max}^k)\,H(W_r-W_0^k(z_i)),
\end{equation}
where $H$ is the Heaviside step function, $[z_{\rm min}^k, z_{\rm max}^k]$ are the minimum and
maximum redshifts observed for the $k$th QSO in Table 1, and the sum extends over all QSO sightlines
in the echelle sample.  The total redshift path length covered by the survey for a given equivalent
width threshold $W_r$ is therefore defined by
\begin{equation}
\Delta\,z\,(W_r)\equiv g(W_r)=\int g(W_r,z)\,dz.
\end{equation} 
We evaluate equations (1) and (2) based on a redshift interval of $\delta\,z=0.0001$.  The results
are shown in panel (a) of Figure 3 for three different equivalent width thresholds, $W_r=30, 50, 80$
m\AA.  The plot illustrates that, as stronger absorber are considered, the total survey path length
increases; and that weaker absorbers require higher signal-to-noise data.  Several strong dips are
immediately apparent at redshifts $z \approx 0.18, 0.21, 0.22, 0.26, 0.29 $, which correspond to
Galactic absorption from \Lya\,1215, \SII\,1250, \SII\,1259, \OI\,1302, \CII\,1334 respectively.

\begin{figure*}
  \epsscale{1.0} 
  \plotone{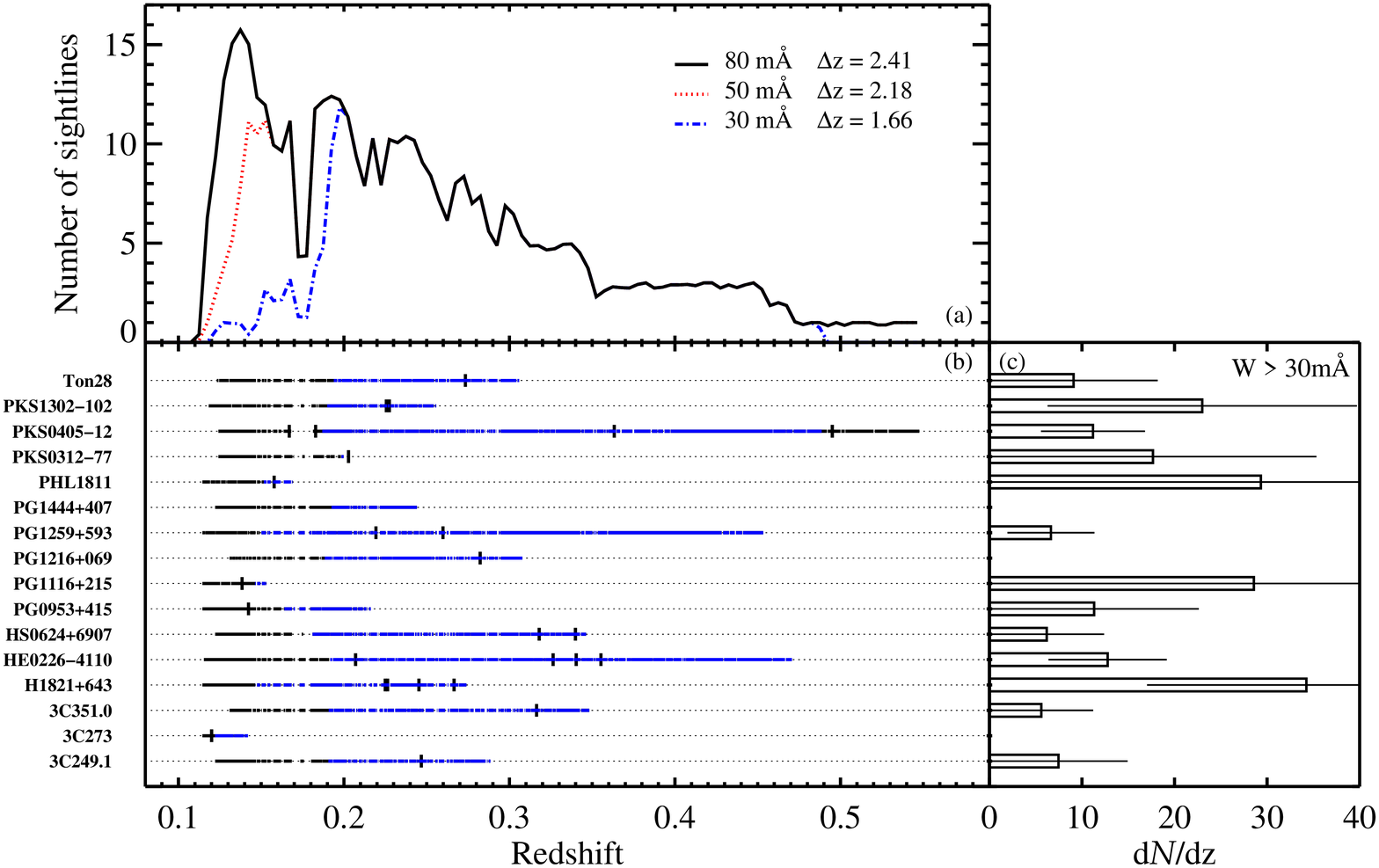}
  \caption{Summary of the results from our blind search of O\,VI
    absorbers.  In Panel (a), we present the total redshift path
    covered by the entire sample of 16 sightlines for different
    absorption equivalent width thresholds $W_*$.  Based on a series
    of Monte-Carlo simulations, we derive a total redshift path of
    $\Delta\,z=1.66, 2.18$, and 2.42 for $W_r=30, 50, 80$ m\AA,
    respectively.  In panel (b), we show for each individual QSO
    sightline the total redshift path allowed by the STIS echelle
    spectrum (black solid lines) and the redshift path that is
    sufficiently sensitive for detecting a $W_r=30$ m\AA\ absorber at
    $> 99.7$\% confidence level (blue lines).  In panel (c), we
    present the number density of $W_r\ge 30\,\mA$ O\,VI absorbers per
    sightline evaluated based on individual lines of sight, together
    with their associated errors.  The errors are estimated according
    to Equation (4).  We see a significant variation across the 16
    sightlines that cannot be easily accounted for by random errors.}
  \label{fig: gz}
\end{figure*}

In panel (b) of Figure 3, we show the redshift path along individual sightlines covered by our
survey (solid black lines).  Masked out regions due to Galactic absorption features (e.g.\ the \lya\
1215 transition) or other ionic transitions of known intervening absorbers appear as gaps in the
path.  Blue segments indicate the spectral regions that are sensitive to $W_r=30$ m\AA.  The
tick-marks in panel (b) indicate the location of random O\,VI absorbers identified in our survey.
We summarize the completeness of our O\,VI search in columns (5) and (6) of Table 2, where we list
the total redshift path sensitive to $W_r=30$ and 50 m\AA\ for each line of sight.

\section{Statistical Properties of the O\,VI Absorbers}
\label{sec: results}

\begin{deluxetable*}{p{1.3in}crccccr}
\tabletypesize{\scriptsize}
\tablecaption{\label{tab: target_summary}Summary of the Blind O\,VI Survey and the Survey Limits}
\tablewidth{0pt}
\tablehead{ & \colhead{} & \colhead{$W_r^{\rm min}$}  & \colhead{Associated} & 
\multicolumn{2}{c}{$\Delta\,z$ ($W\ge W_*$)} & & \colhead{\dndz} \\
\cline{5-6} 
\cline{8-8} 
\colhead{QSO} & \colhead{$N_{\mbox{abs}}^{\rm random}$} & \colhead{$(\mA)$} &  
\colhead{Absorbers} & \colhead{$(W_* = 50\,\mA)$} & 
\colhead{$(W_* = 30\,\mA)$} & & \colhead{$(W_* = 30\,\mA)$} \\
\colhead{(1)} & \colhead{(2)} & \colhead{(3)} & \colhead{(4)} & \colhead{(5)} & 
\colhead{(6)} & & \colhead{(7)} }
\startdata
HE\,0226$-$4110\tablenotemark{1,2,3} \dotfill & 4 &  43 & Y & 0.3055 & 0.2593 & & $13\pm 7$ \\
PKS\,0312$-$77                       \dotfill & 1 & 655 & N & 0.0375 & 0.0014 & & $18\pm 18$ \\
PKS\,0405$-$12\tablenotemark{4,5}    \dotfill & 4 &  30 & N & 0.3542 & 0.2780 & & $12\pm 6$ \\
HS\,0624$+$6907\tablenotemark{6}     \dotfill & 2 &  27 & Y & 0.1681 & 0.1458 & & $6\pm 6$ \\
PG\,0953$+$415\tablenotemark{7,8}    \dotfill & 1 & 121 & Y & 0.0698 & 0.0434 & & $11\pm 11$ \\
3C\,249.1                            \dotfill & 1 &  72 & Y & 0.1206 & 0.0826 & & $8\pm 8$ \\
PG\,1116$+$215\tablenotemark{9,10}   \dotfill & 1 &  76 & Y & 0.0307 & 0.0053 & & $28\pm 28$ \\
PG\,1216$+$069                       \dotfill & 1 &  26 & N & 0.1339 & 0.1047 & & $0\pm 10$ \\
3C\,273\tablenotemark{11}            \dotfill & 1 &  22 & N & 0.0271 & 0.0195 & & $0\pm 51$ \\
PG\,1259$+$593\tablenotemark{12}     \dotfill & 2 &  77 & N & 0.2851 & 0.2680 & & $7\pm 5$ \\
PKS\,1302$-$102                      \dotfill & 2 &  30 & N & 0.0992 & 0.0596 & & $26\pm 19$ \\
PG\,1444$+$407                       \dotfill & 0 & ... & Y & 0.0927 & 0.0478 & & $0\pm 21$ \\
3C\,351.0\tablenotemark{13}          \dotfill & 1 & 232 & Y & 0.1575 & 0.1313 & & $6\pm 6$ \\
PHL\,1811\tablenotemark{14}          \dotfill & 1 &  64 & Y & 0.0388 & 0.0133 & & $23\pm 23$ \\
H\,1821$+$643\tablenotemark{15}      \dotfill & 4 &  30 & Y & 0.1226 & 0.0999 & & $34\pm 17$ \\
Ton\,28                              \dotfill & 1 &  26 & Y & 0.1396 & 0.0991 & & $0 \pm 10$\\
\hline 
total (16 sightlines)                         &27 &     & 10&  2.18  &  1.66  & & $10\pm 2$       \\
\enddata
\tablecomments{See also: (1) \citet{lehner-etal-06-HE0226-4110}; (2) \citet{savage-etal-05-HE0226-4110-NeVIII}; (3) \citet{ganguly-etal-06-HE0226-4110_qso_host_lines}; (4) \citet{chen-prochaska-00-PKS0405-12_z0.167}; (5) \citet{prochaska-etal-04-PKS0405-12}; (6) \citet{aracil-etal-06a-metals}; (7) \citet{tripp-savage-00-PG0953+415_z0.1423}; (8) \citet{savage-etal-02-PG0953+415}; (9) \citet{sembach-etal-04-PG1116+215}; (10) \citet{richter-etal-06-BLAs}; (11) \citet{heap-etal-02}; (12) \citet{richter-etal-04-PG1259+593}; (13) \citet{yuan-etal-02-3C351-qso-lines}; (14) \citet{jenkins-etal-03}; (15) \citet{tripp-etal-00-H1821+643}.}
\end{deluxetable*}

We detect a total of \nabs\ random \OVI\ doublets along 16 QSO lines of sight over a redshift range
from $z_{\rm O\,VI}=0.12$ to $z_{\rm O\,VI}=0.495$ with a mean redshift of $\langle z\rangle =
0.25$.  Furthermore, we uncover associated absorbers toward nine QSOs at $|\Delta\,v|< 5000$ \kms\
from the background QSOs.  The random O\,VI absorbers form a statistically representative sample for
constraining their global mean properties.  The associated O\,VI absorbers form a uniform sample for
studying the gaseous halos of distant QSO hosts.  Here we consider the statistical properties of the
two samples separately.  Studies of the physical nature of individual absorbers will be presented in
a separate paper (Thom \& Chen 2008; Paper II).

\subsection{The Incidence of Random O\,VI Absorbers}

The Monte-Carlo simulations presented in \S\ 3.2 allow us to estimate the total redshift path length
as a function of absorption equivalent width, from which we can then derive an accurate estimate of
the number density of O\,VI absorbers per line of sight, using individual sightlines separately or
the entire QSO echelle sample collectively.  For each absorber of rest-frame O\,VI 1031 absorption
equivalent width $W_r^i>W_*$ identified in our search, the contribution to the total number density
per line of sight $d{\cal N}/dz$ is $1/g(W_r^i)$.  Over a statistical sample, we can therefore
derive for a given equivalent width limit $W_*$,
\begin{equation}
d\,{\cal N}(W \ge W_*)/dz = \sum_{i}\,\frac{1}{\Delta\,z (W_r^i)}\,\times\,H(W_r^i-W_*)
\end{equation}
where the sum extends over all systems found in a random sample and $H$ is the Heaviside step
function.  Note that $\Delta\,z\,(W_r^i)$ varies with $W_r^i$.  A stronger absorber may be detected
in lower $S/N$ spectra, allowing a larger survey path length for a given QSO echelle sample.
Equation (3) is therefore not the total number of absorbers, $N_{\rm abs}^{\rm random}$, divided by
a fixed path length, $(\Delta\,z)$.  The variance of $\dndz$ is evaluated according to
\begin{equation}
\sigma_{d{\cal N}/dz}^2 = \sum_{i}\,\frac{1}{[\Delta\,z (W_r^i)]^2}\,\times\,H(W_r^i-W_*).
\end{equation}

We first evaluate \dndz\ separately for individual lines of sight, adopting a common threshold for
all QSO spectra.  The results for $W_*=30$ \mA\ are summarized in column (7) of Table 2, and
presented in panel (c) of Fig.~\ref{fig: gz}.  Despite the relatively short path length per line of
sight, it is immediately apparent from Fig.~\ref{fig: gz} that there is a substantial variation in
\dndz\ from one line-of-sight to another.  In contrast to the line-of-sight toward H\,1821$+$643,
along which four O\,VI absorbers are found over $\Delta\,z\lesssim 0.1$ \citep{tripp-etal-00-H1821+643},
we note that the sightlines toward HS\,0624$+$6907, PG1259$+$593, and 3C\,351.0 allow longer survey
paths but exhibit fewer absorbers (between 1 and 2).  The large scatter in \dndz\ between random
sightlines underscores the necessity of a statistically significant sample of O\,VI absorbers for an
unbiased estimate of their incidence.

Combining all 16 QSO lines of sight together, we derive $d\,{\cal N}(W\ge 50\,\mA)/dz = 6 \pm 3$
over $\Delta\,z\,(W_*=50\,\mA)=2.18$, and $d\,{\cal N}(W\ge 30\,\mA)/dz = 9 \pm 4$ over
$\Delta\,z\,(W_*=30\,\mA)=1.66$.  In \S\ 5 and panel (b) of Figure 5, we  show that our number
density measurements can be explained well by recent numerical simulations that do not require super
starburst outflow, but account for non-equilibrium ionization conditions \citep{cen-fang-06}.

\subsection{The Cosmological Mass Density of Ionized Gas}

A particularly interesting quantity to determine, based on a statistically representative O\,VI
absorber sample, is the cosmological mean mass density contained in the O$^{5+}$ ions, $\Omega_{{\rm
    O}^{5+}}$.  In principle, we can then derive the cosmological mean mass density of warm baryons,
$\Omega_{\rm b}$, probed by the O\,VI transition according to

\begin{equation}
\Omega_{\rm b}=\frac{1}{f}\,\frac{\mu}{Z\times ({\rm O}/{\rm H})_\odot}\,\frac{m_{\rm H}}{m_{\rm O}}\,\Omega_{{\rm O}^{5+}},
\end{equation}
where $f\equiv N({\rm O}^{5+})/N({\rm O})$ is the fiducial ionization fraction of the O\,VI
absorbing gas, $Z\equiv ({\rm O}/{\rm H})/({\rm O}/{\rm H})_\odot$ represents its mean metallicity,
$({\rm O}/{\rm H})_\odot$ is the solar oxygen to hydrogen ratio by number, $\mu$ is the mean atomic
mass of the gas ($\mu=1.3$ for gas of solar composition), and $m_{\rm H}$ and $m_{\rm O}$ are the
atomic weight of hydrogen and oxygen, respectively.  In the following calculation, we adopt $({\rm
  O}/{\rm H})_\odot=4.57\times 10^{-4}$ (Asplund \etal\ 2004).

Following \citet{lanzetta-etal-01}, we calculate $\Omega_{{\rm O}^{5+}}$ according to
\begin{eqnarray}
\Omega_{{\rm O}^{5+}} &=& \frac{H_0\,m_{\rm O}}{c\,\rho_c} \int N({\rm O\,VI})\,f(N({\rm O\,VI}))\,d\,N({\rm O\,VI}) \\
&=& \frac{H_0\,m_{\rm O}}{c\,\rho_c} \sum_i\,\frac{N_i({\rm O\,VI})}{\Delta\,X_i}
\end{eqnarray}
where $\rho_c$ is the present-day critical density, $f(N({\rm O\,VI}))$ is the frequency distribution
function defined as the number of O\,VI absorbers per column density interval per unit absorption
path-length, $N_i$ is the corresponding O\,VI absorption column density for absorber $i$ with
rest-frame absorption equivalent width $W_r^i$ and $\Delta\,X_i$ is the co-moving absorption path
length covered by the entire echelle sample at the absorption strength $W_r^i$.  The co-moving
absorption path length $\Delta\,X_i$ for absorber $i$ is related to the total redshift path length
$\Delta\,z$ evaluated from Equation (2) at $W_r^i$ according to
\begin{equation}
\frac{d\,X}{d\,z} = \frac{(1+z)^2}{[\Omega_M\,(1+z)^3 - (\Omega_M +
\Omega_{\Lambda} - 1)(1+z)^2 + \Omega_{\Lambda}]^{1/2}}.
\end{equation}
The corresponding 1-$\sigma$ error is evaluated according to
\begin{equation}
\sigma_{\Omega_{{\rm O}^{5+}}} = \frac{H_0\,m_{\rm O}}{c\,\rho_c} \left[\sum_i\,\frac{\sigma_{N_i({\rm O\,VI})}^2}{(\Delta\,X_i)^2}\right]^{1/2}.
\end{equation}
For each uncovered absorber of $W_r^i$, we perform a Voigt profile fit to determine the
corresponding $N_i$.

We present the frequency distribution function determined based on the random O\,VI absorber sample
in panel (a) of Figure 5.  The bin size is selected to contain roughly the same number of absorbers
per bin.  Error bars represent 1-$\sigma$ Poisson fluctuations.  For the sample of 23 random O\,VI
absorbers with $W_r > 30\mA$, we therefore derive $\Omega({\rm O}^{5+})\,h = \OmegaOVInum$.
Assuming a $Z=1/10\,Z_\odot$ mean metallicity and $f=0.2$ ionization fraction for the O\,VI gas, we
further derive $\Omega_{\rm b}h\approx 0.0014$.  Bearing in mind the large uncertainties in the
metallicity and ionization fraction of the O$^{5+}$ bearing gas, our derived mean mass density of
ionized gas is less than 10\% of the total baryon mass density (e.g.\ Burles \etal\ 2001).

\subsection{Associated O\,VI Absorbers}

\begin{figure*}
  \epsscale{1.0} 
  \plotone{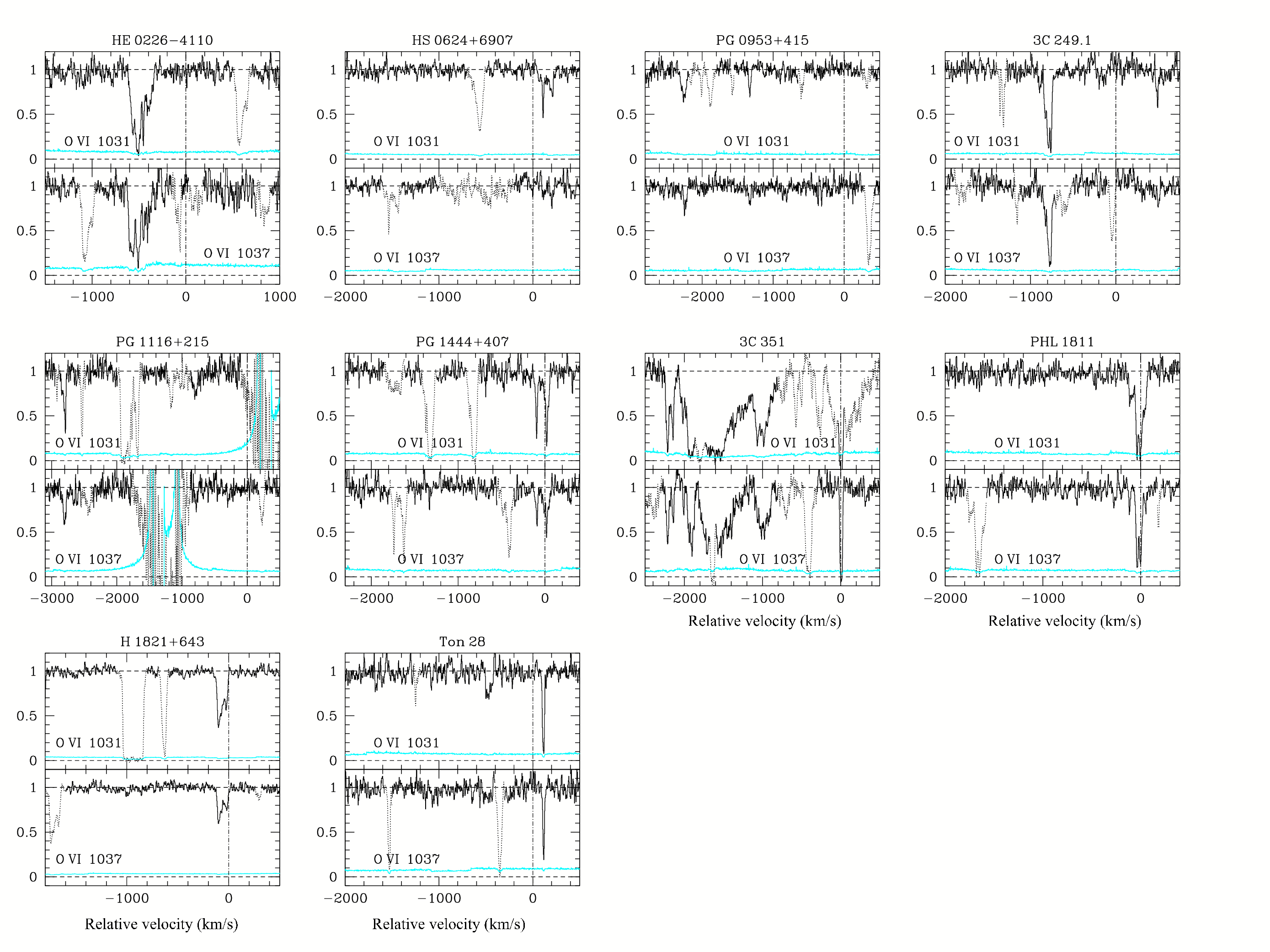}
  \caption{Absorption profiles of associated O\,VI absorbers
  identified in our STIS echelle sample.}
  \label{fig: qso}
\end{figure*}

We have excluded spectral regions within $|\Delta\,v|=0-5000$ \kms\ from the background QSOs for
establishing the statistical sample of random foreground absorbers in the previous discussion, due
to potential contamination features from the QSO host environment.  However, the blind O\,VI search
has been carried out all the way through the emission redshifts of these QSOs.  The O\,VI absorbers
found in the proximity of QSOs form a sample of associated absorbers that offers important
constraints on the halo gas in QSO hosts (Chelouche \etal\ 2007, in prep).

Ten of the 16 QSOs in our STIS echelle sample exhibit strong O\,VI absorbers $|\Delta\,v|\le 5000$
\kms\ blue-shifted from their emission redshifts (indicated in Table 2).  We present the absorption
profiles of these O\,VI absorbers in Figure 4.  The rest-frame absorption equivalent width of the
1031 member ranges from $W_r=60\ \mA$ (the absorber at $\Delta\,v=-480$ \kms\ from Ton\,28) to
$W_r=3$ \AA\ (the multi-component absorber at $\Delta\,v\approx -1600$ \kms\ from 3C\,351.0).  The
multi-component absorber found in the vicinity of 3C\,351.0 has complex velocity profiles, extending
from $\Delta\,v\approx 0$ \kms\ to $\Delta\,v\approx -2200$ \kms.  It exhibits evidence of partial
covering of the O$^{5+}$ ions and is understood to originate in the outflow gas from the background
QSO \citep{yuan-etal-02-3C351-qso-lines}.

Excluding the complex sightline toward 3C351.0, the total redshift path length defined within
$|\Delta\,v|=0-5000$ \kms\ blue-ward of the emission redshifts of the remaining 15 QSOs in our
echelle sample is $\Delta\,z=0.33$.  At the equivalent width limit of our search $W_*=50\ \mA$, we
found nine O\,VI absorbers.  We measure $d\,{\cal N}/dz=27\pm 9$ for O\,VI absorbers of
$W_r(1031)>50\ \mA$ in the vicinity of background QSOs.  This is significantly higher than the
incidence of random absorbers in the foreground.

\section{Comparisons with Previous Surveys}

Our search in the STIS echelle QSO sample has uncovered a total of \nabs\ random \OVI\ absorbers of
$W_r\gtrsim 25\,\mA$ along 16 lines of sight, from which we derive the incidence of random O\,VI
absorbers, $d\,{\cal N}(W\ge 30\,\mA)/dz = $ \dndznum\ at $\langle z\rangle = 0.25$.  This is roughly
half of what is mentioned in Tripp \etal\ (2006), who reported a total of 44 O\,VI absorbers along
16 lines of sight and $d\,{\cal N}(W\ge 30\,\mA)/dz = 23 \pm 4$ at $z<0.5$.  An independent survey
of O\,VI absorbers in known \lya\ absorbers by \citet{danforth-etal-05-OVI-baryon-census} yields an
absorber frequency of $d\,{\cal N}(W\ge 30\,\mA)/dz = 17 \pm 3$ or $d\,{\cal N}(W\ge 50\,\mA)/dz = 9
\pm 2$ at $z<0.15$.  We present these different measurements in Figure 5, along with recent
numerical simulation results that include galactic wind feedback and non-equilibrium ionization
conditions \citep{cen-fang-06}.  It is interesting to note that models with starburst feedback are
not required by our measurements.

The results of our blind search are consistent with those of
\citet{danforth-etal-05-OVI-baryon-census} from a FUSE survey at lower redshift that had been
carried out in known \lya\ absorbers.  The agreement is qualitatively consistent with the small
fraction of O\,VI absorbers that do not have an associated \lya\ transition.  The discrepancy
between our results and those of \citet{tripp-etal-06-conf} based on a similar set of QSO spectral
sample is most puzzling.  In the following, we discuss possible reasons for this discrepancy.

It is possible that the discrepancy is due to different QSO spectra covered by different surveys.
Our random sample has been established based exclusively on available STIS echelle data.  The
sightline along H\,1821$+$643 has the highest incidence of O\,VI absorbers of all 16 sightlines in
our echelle sample (See Table 2 and the discussion in \S\ 4.1).  In addition to the four O\,VI
absorbers originally reported in \citet{tripp-etal-00-H1821+643} and confirmed in our blind search,
\citet{tripp-etal-01-H1821+643_z0.1212} reported an additional O\,VI absorber at $z=0.1212$ in a
targeted search in higher quality FUSE spectra.  At $z=0.1212$, the O\,VI $\lambda\,1031$ line is
redshifted to $\lambda_{\rm obs}\approx 1157$ \AA, where the sensitivity of STIS echelle declines
steeply.  The lack of sensitivity at the blue end of the echelle sample is accounted for by the
completeness study of our random sample discussed in \S\ 3.2.  This $z=0.1212$ absorber found in a
targeted search after including additional FUSE spectra does not satisfy the criterion for
establishing a statistically unbiased sample, and has therefore been excluded in our random sample.

In addition, there are O\,VI absorbers reported in the literature but not confirmed by our search.
Along the same redshift path-length covered by our survey, 16 foreground O\,VI systems have been
reported in the literature by other authors toward seven QSOs (HE\,0226$-$4110, PKS\,0405$-$123,
PG\,0953$+$415, PG\,1116$+$215, PG1259$+$593, H\,1821$+$643, and 3C\,273; See Table 2 for
corresponding references), four are not confirmed in our survey.  An example is shown in panel (d)
of Figure 1, where we show the spectral region of an O\,VI doublet along the sightline toward
PG\,1259$+$593 reported by \citet{richter-etal-04-PG1259+593}.  The putative O\,VI 1031 member is
blended with the Si\,III 1206 transition of a \lya\ absorber at $z=0.0461$ (the feature at $v=-30$
\kms) and the C\,III transition of a \lya\ absorbers at $z=0.2924$ (the feature at $v=+100$ \kms).
Our search routine did not accept this absorber due to discrepant line strengths.  Our random sample
does not include this possible absorber.  Along the same sightline,
\citet{richter-etal-04-PG1259+593} also reported the presence of an O\,VI absorber at $z=0.3198$,
which is not confirmed by our routine due to discrepant absorption profiles and an inconsistent line
ratio.

Along the sightline toward HE\,0226$-$4110, \citet{lehner-etal-06-HE0226-4110} reported five O\,VI
absorbers with $W_r(1031)>50\,\mA$.  We cannot confirm the presence of their identification for a
$z=0.4266$ absorber.  In particular, the putative feature identified as the O\,VI 1037 member is not
present in our own reduction of the echelle spectra.  Inspections of individual exposures showed
that the broad feature are predominant in two of the 12 exposures, but the discrepancy may be due to
different treatment in the hot pixels in echelle spectra (T.\ Tripp, private communication).
Additional spectra obtained using the Cosmic Origin Spectrograph (COS) or STIS will be valuable for
confirming the presence of the doublet.  Consequently, this absorber is not included in our random
sample.

\section{Discussion}
\label{sec: discussion}

\subsection{The Ionization State of the O\,VI Gas}

\begin{figure}[t]
 \epsscale{1.0}
 \plotone{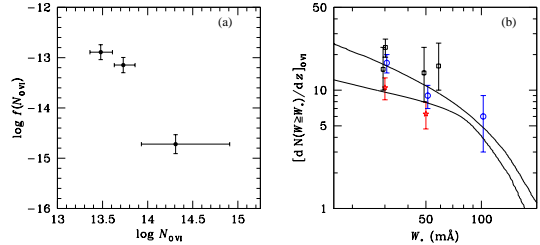}
 \caption{(a) The frequency distribution function of O\,VI absorbers
   from our random sample.  The bin size is chosen for each bin to
   contain roughly the same number of absorbers.  Error bars represents
   1-$\sigma$ Poisson fluctuations.  (b) Observed incidence of O\,VI
   absorbers from different studies. Star points are from this work;
   open circles are from \citet{danforth-etal-05-OVI-baryon-census} ;
   and open squares are from \citet{savage-etal-02-PG0953+415,
   prochaska-etal-04-PKS0405-12, tripp-etal-06-conf, cooksey-etal-08}.
   Note that we have slightly offset the data points in the horizontal
   direction for clarification.  The curves are recent WHIM
   predictions based on numerical simulations \citep{cen-fang-06} that
   account for non-equilibrium ionization conditions.  The top curve
   further includes Galactic wind feedback.}
 \label{fig: dndz}
\end{figure}

Various studies of individual O\,VI absorbers that consider both kinematic signatures of the
absorption profiles and abundance ratios between different ions have concluded that a large fraction
of these absorbers can be explained by a simple photo-ionization model
\citep[e.g.][]{tripp-etal-00-H1821+643, prochaska-etal-04-PKS0405-12, sembach-etal-04-PG1116+215,
  lehner-etal-06-HE0226-4110}.  These results are in contradiction to the expectation of a WHIM
origin, under which the bulk of the O$^{5+}$ ions originate in the shocks created as the gas
accretes onto the large filamentary structures.  However, it is often difficult to rule out a
collisional ionization scenario for these absorbers.  Additional uncertainties arise, if O\,VI
absorbers arise in post-shock gas that is undergoing radiative cooling and a simple ionization
equilibrium assumption does not apply (Gnat \& Sternberg 2007).

The comparison of the absorption column densities in Figure 6 between \lya\ and O\,VI transitions
for the O\,VI selected sample (solid points) demonstrates that the $N(\hI)$ associated with these
O\,VI absorbers spans nearly five orders of magnitude.  In addition, stronger O\,VI transitions have
on average stronger associated \lya\ transitions.  The null hypothesis in which \lya\ and O\,VI
column densities are randomly distributed among each other can be ruled out at $>97$\% confidence
level based on a generalized Kendall test that considers the non-detection of \lya\ for the O\,VI
absorber at $z=0.32639$ toward HE\,0226$-$4110.  We note that the STIS echelle sample is sensitive
to O\,VI absorbers of $W_r>80$ \mA\ over the entire redshift path-length allowed by the data, and
O\,VI absorbers of $W_r>50$ \mA\ over $\approx 80$\% of the sightlines.  The echelle spectra
therefore allow us to uncover O\,VI absorbers of $\log\,N({\rm O\,VI}) > 13.5$ over the majority of
the sightlines.  The correlation is therefore unlikely due to a selection bias.  Including
measurements from known O\,VI absorbers at low redshift (open squares and crosses), we find that the
correlation remains at $>80$\% significance level.

The broad range of $N(\hI)$ spanned by these O\,VI absorbers also makes it difficult to apply a
simple collisional ionization equilibrium model for explaining their origin.  Furthermore, the
correlation between $N({\rm O\,VI})$ and $N(\hI)$ indicates that the O\,VI absorbers appears to
track closely H\,I gas in overdense regions, while the opposite does not apply
\citep[e.g.][]{danforth-etal-05-OVI-baryon-census}.  This trend can be understood under a simple
photo-ionization scenario, and the slope in $N({\rm O\,VI})$ versus $N(\hI)$ implies a declining
ionization parameter (defined as the number of ionizing photons per gas particle) with increasing
gas column density (see Figure 31 of Paper II).

\subsection{O\,VI Absorbers without Associated H\,I}

A unique feature of our blind search is that the finding of O\,VI doublet does not rely on known
presence of \lya\ transitions.  The sample therefore allows us to constrain the fraction of O\,VI
absorbers that do not show significant traces of H\,I.  These absorbers may represent an extreme
case for outflow gas that has been ejected into underdense regions by strong galactic winds
\citep[e.g.][]{kawata-rauch-07}.

In Figure 6, we find that 26 of the 27 O\,VI absorbers in the random sample show associated H\,I
absorption features.  The only one possible \lya-free system is found to have $N({\rm
  O\,VI})=13.6\pm 0.1$ at $z_{\rm O\,VI}=0.3264$ toward HE\,0226$-$4110 (the arrow in Figure 6).
The comparison indicates that $<5$\% of O\,VI absorbers originate in underdense regions of
$N(\hI)\lesssim 10^{13}$ \cmjj\ that are heavily metal-enriched by starburst outflows

\subsection{Large-scale Variation in Absorber Number Density}

\begin{figure*}
\epsscale{0.6}
\plotone{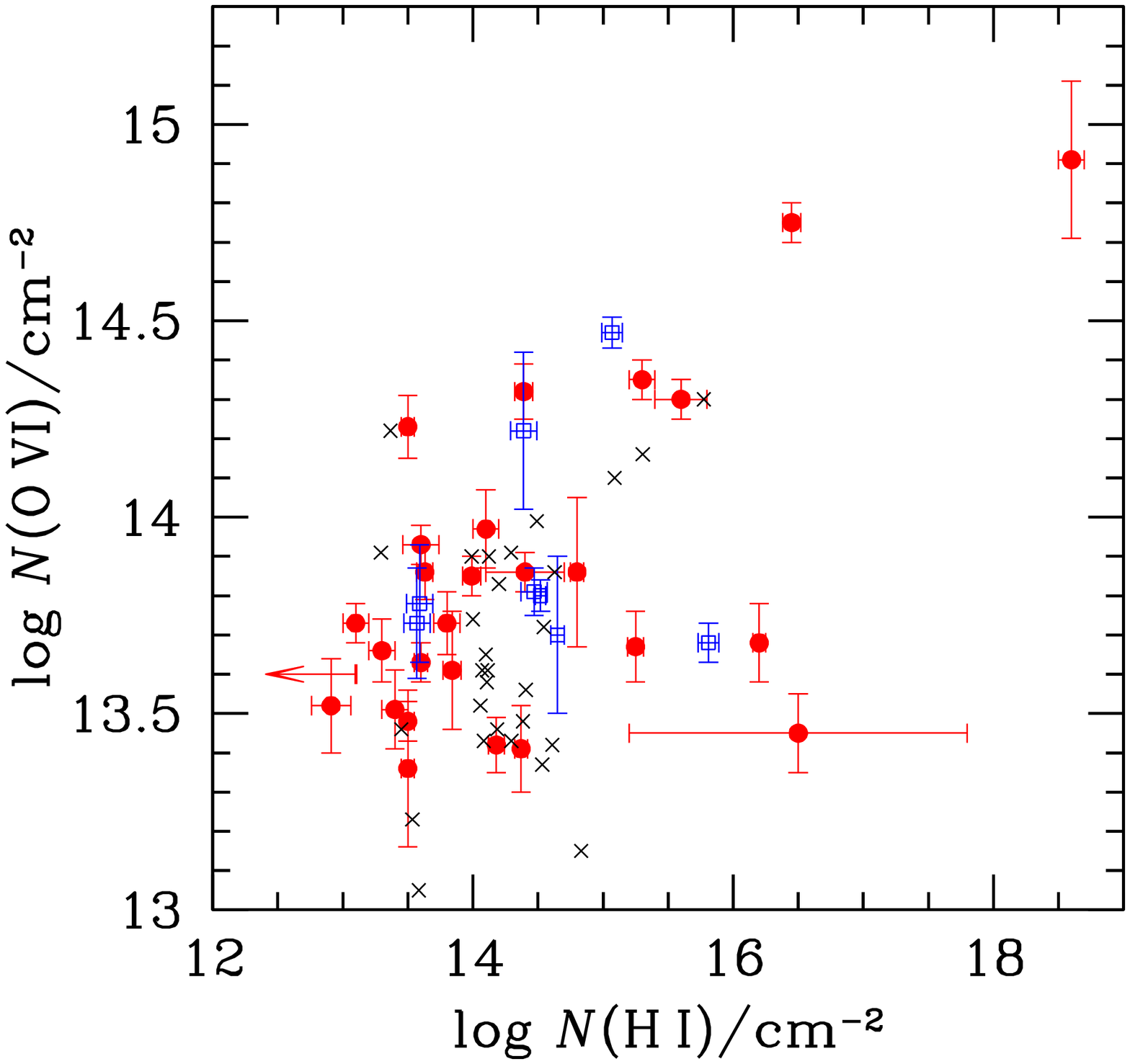}
\caption{Comparison between $N({\rm O\,VI})$ and $N(\hI)$ for the 27
  O\,VI absorbers identified in our blind search (solid points.  All
  absorbers but one have an associated \lya\ absorption line.  We have
  also included published absorbers at low redshifts.  Open squares
  are $z<0.12$ absorbers presented in
  \citet{savage-etal-02-PG0953+415} for PG\,0953$+$415, in
  \citet{sembach-etal-04-PG1116+215} for PG\,1116$+$215, in
  \citet{richter-etal-04-PG1259+593} fo PG\,1259$+$593, in
  \citet{prochaska-etal-04-PKS0405-12} for PKS\,0405$-$123, and in
  \citet{cooksey-etal-08} for PKS\,1302$-$102.  Crosses indicate
  measurements from a FUSE survey by
  \citet{danforth-etal-05-OVI-baryon-census}.}
  \label{fig: ncomp}
\end{figure*}

Cosmological simulations also show that O\,VI absorbers originating in the WHIM reside in regions of
over-density $\delta\equiv \rho/\langle\rho\rangle=10-100$ at $z\sim 0$
\citep{dave-etal-01-WHIM-baryons, cen-etal-01, fang-bryan-01-H1821+643_IGM}.  This is similar to
what is found in simulation studies of \lya\ forest absorbers, according to which by $z=0$ the
majority of \lya\ absorbers are associated with $\delta = 10-100$.  It is therefore interesting to
examine the relation between \lya\ and O\,VI absorbers.

Of the 16 QSO sightlines in our echelle sample, seven have been studied extensively for the
line-of-sight properties of \lya\ absorbers \citep{sembach-etal-04-PG1116+215, williger-etal-06,
  lehner-etal-06-HE0226-4110}.  We present in Figure 7 a comparison of the observed number densities
between \lya\ absorbers and O\,VI absorbers measured along individual sightlines.  These include
H\,1821$+$643, HE\,0226$-$4110, HS\,0624$+$6907, PG\,0953$+$415, PG\,1116$+$215, PG\,1259$+$593, and
PKS\,0405$-$123.  The statistics of \lya\ absorbers are taken from the compilation of
\citet{lehner-etal-06-HE0226-4110}.  We have included the corrected \lya\ absorber catalog for the
PKS\,0405$-$123 sightline.  The error bars represent 1-$\sigma$ uncertainties.

Considering all seven sightlines together, we find a mean number density of $\langle\,d\,{\cal
  N}/dz\rangle=66$ for the \lya\ absorbers with an r.m.s.\ dispersion of 15.  In contrast, we find
$\langle\,d\,{\cal N}/dz\rangle=12$ for the O\,VI absorbers with an r.m.s.\ dispersion of 10.
Examining the number density variation from one sightline to another, we find that the line of sight
toward H\,1821$+$643 with the highest incidence of O\,VI absorbers has the lowest incidence of \lya\
absorbers.  Figure 7 shows a possible inverse correlation between the number density of \hI\
absorbers and the number density of O\,VI absorbers.  While \lya\ absorbers appear to be relatively
more abundant than O\,VI along each sightline, the number density of O\,VI absorbers do not scale
proportionally with the number density of \lya\ absorbers along individual lines of sight.

This anti-correlation between $[d{\cal N}/dz]_{\rm O\,VI}$ and $[d{\cal N}/dz]_{\rm \lya}$ may be
understood, if those sightlines with a higher incidence of O\,VI absorbers probe a larger fraction
of more highly ionized regions.  However, we note the result from \S\ 6.1, where we find nearly
every one of the O\,VI absorbers in our statistical sample shows an associated \lya\ transition.  On
the other hand, if O\,VI absorbers originate preferentially in underdense regions, then a higher
incidence of O\,VI implies a substantial fraction of the sightline probe underdense regions and
hence a lower incidence of \lya\ absorbers.  This is expected if heavy elements from starburst
outflows are ejected primarily into underdense regions \citep[e.g.][]{kawata-rauch-07}.  Further
insights can be obtained from a detailed examination of the large-scale galaxy environment of the
O\,VI absorbers.  A larger sample of QSO echelle spectra will also be valuable for a more
quantitative study to confirm this result.

\begin{figure*}
\epsscale{0.6}
\plotone{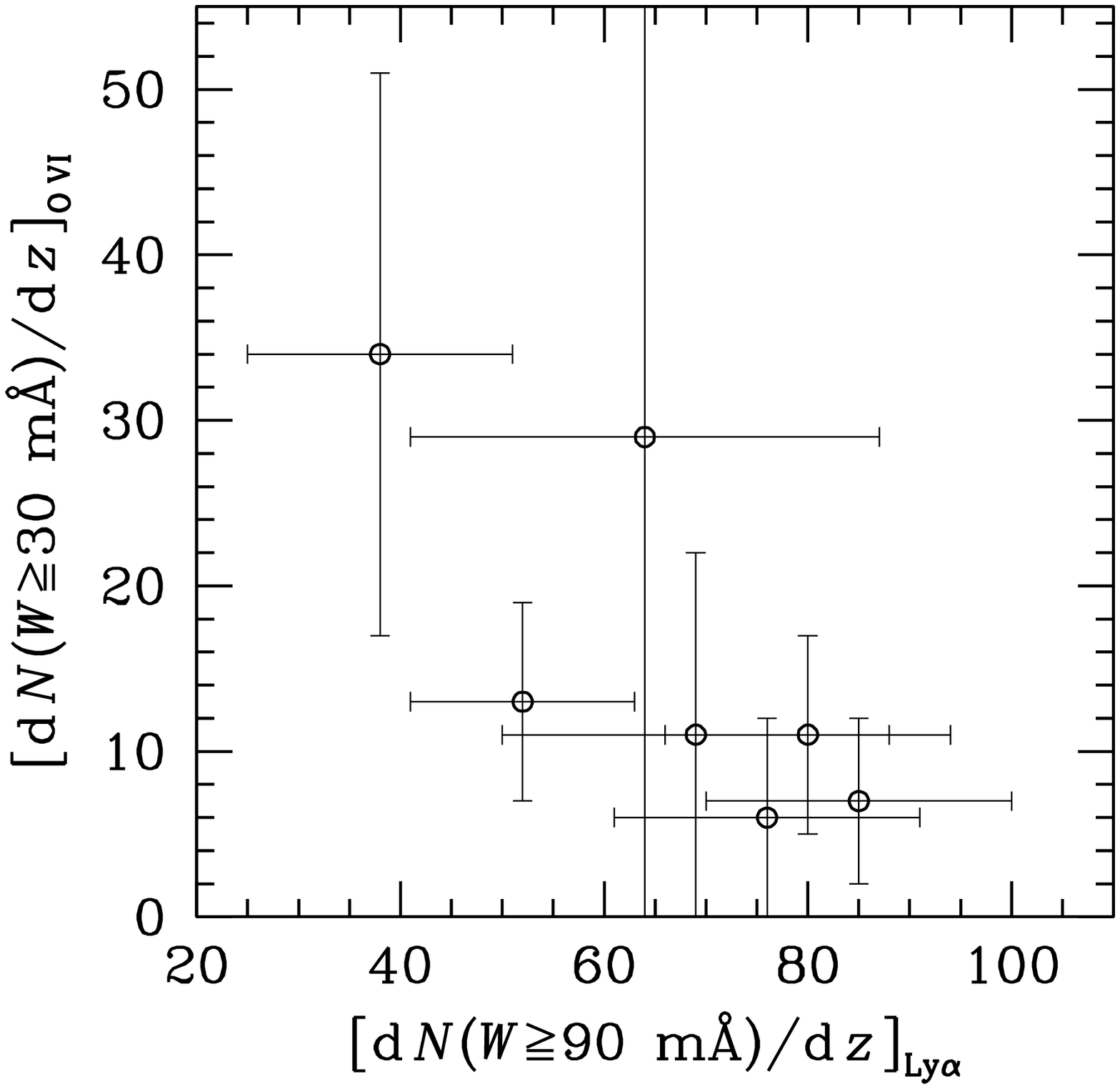}
\caption{Comparison of the line density variation from one line of
  sight to another for hydrogen \lya\ absorbers and O\,VI absorbers.
  The number density of \lya\ absorbers measured for individual
  sightlines are taken from Lehner \etal\ (2007) for \lya\ absorbers
  of rest-frame equivalent width $W_r(\lya) > 90 \mA$ and the number
  density of O\,VI absorbers measured for individual sightlines are
  from the blind search for O\,VI absorbers of rest-frame equivalent
  width $W_r({\rm O\,VI}) > 30 \mA$ presented in this paper.  We find a
  possible inverse correlation between the number density of \hI\
  absorbers and the number density of O\,VI absorbers.}
  \label{fig: lden}
\end{figure*}

\section{Summary}

We have conducted a systematic survey of intervening O\,VI absorbers in available echelle spectra
obtained using HST/STIS with the E140M grating.  We have collected echelle spectra of 16 distant
QSOs from the HST data archive.  These UV bright quasars at $z_{\rm QSO}=0.17-0.57$ have been
selected for HST/STIS observations by different authors for different studies, and represent a
random, flux-limited sample of QSO echelle spectra for an unbiased search of intervening absorbers.
The primary objectives of our blind O\,VI survey are: (1) to obtain a statistically representative
sample of O\,VI absorbers along random lines of sight for an accurate measurement of the
cosmological mass density of warm-hot gas at low redshift; and (2) to study the incidence of O\,VI
absorbers along individual sightlines for quantifying the spatial variation of O\,VI absorbing gas.

Our search in the STIS echelle QSO sample has uncovered a total of \nabs\ random ($|\Delta\,v|>5000$
\kms\ from the background QSO) \OVI\ absorbers with rest-frame absorption equivalent width of the
1031 member $W_r\gtrsim 25\,\mA$ along 16 lines of sight.  Ten of these QSOs also exhibit strong
O\,VI absorbers at $|\Delta\,v|<5000$ \kms\ blue-shifted from the QSO emission redshifts.  Our O\,VI
survey differs from previous studies in two fundamental ways.  First, we do not require prior
knowledge of the presence of other transitions such as \lya.  The echelle resolution allows us to
identify the O\,VI absorption doublet based on their common line centroid and known flux ratio.
Second, the survey completeness is well understood and characterized based on a series of
Monte-Carlo simulations.  We obtain an equivalent width threshold curve for every QSO sightline.
Together, these curves allow us to estimate the total redshift survey path as a function of
equivalent width threshold, thereby maximizing the survey efficiency at the limit allowed by the
echelle spectra.

Based on the results of the Monte-Carlo simulations, we derive a total redshift path for random
absorbers of $\Delta\,z=1.66, 2.18$, and 2.42 for $W_r=30, 50$ and $80$ m\AA, respectively.  The total
path-length surveyed in our sample is comparable to the FUSE survey carried out at $z<0.15$ by
Danforth \& Shull (2005), and is the largest at $z=0.12-0.5$.  The estimated total redshift path
leads to a number density of $d\,{\cal N}(W\ge 50\,\mA)/dz = 6.7 \pm 1.7$ and $d\,{\cal N}(W\ge
30\,\mA)/dz = $ \dndznum.  Recent numerical simulations that allow non-equilibrium ionization
conditions explain the observations very well.  In contrast, we also measure $d\,{\cal N}/dz=27\pm
9$ for O\,VI absorbers of $W_r(1031)>50\ \mA$ in the vicinity of background QSOs.  For the sample of
27 random O\,VI absorbers, we further derive a mean cosmological mass density of the O\,VI ions
$\Omega({\rm O}^{5+})\,h = \OmegaOVInum$ that leads to a mean cosmological mass density of warm
baryons $\Omega_{\rm b}h\approx 0.0014$.

Using the statistical sample of O\,VI absorbers, we show that $<5$\% of O\,VI absorbers originate in
underdense regions that do not show significant trace of H\,I.  In addition, a comparison of the
absorption column densities between \lya\ and O\,VI transitions for the O\,VI selected sample shows
that the $N(\hI)$ associated with these O\,VI absorbers spans nearly five orders of magnitude,
making it difficult to apply a simple collisional ionization equilibrium model for explaining their
origin.  Furthermore, stronger O\,VI transitions have on average stronger associated \lya\
transitions, indicating that the O\,VI absorbers appear to track closely H\,I gas in overdense
regions, while the opposite does not apply.  At the same time, comparisons of individual
line-of-sight properties show a moderate inverse correlation between the number density of \hI\
absorbers and the number density of O\,VI absorbers.  While nearly all O\,VI absorbers have an
associated \lya\ transitions to $N(\hI)\gtrsim 10^{13}$ \cmjj, the presence of abundant O\,VI
absorbers implies that a larger fraction of the gas along the line of sight becomes more ionized and
therefore reducing the line density of \lya\ absorbers.  A larger sample of QSO echelle spectra is
necessary for a more detailed study.

\acknowledgments

We thank N.\ Gnedin, J.~X.~Prochaska, and J.\ Tinker for valuable discussions.  This research has
made use of the NASA/IPAC Extragalactic Database (NED) which is operated by the Jet Propulsion
Laboratory, California Institute of Technology, under contract with the National Aeronautics and
Space Administration.  C.~T. and H.-W.C.\ acknowledges support from NASA grant NNG06GC36G.
H.-W.C. acknowledges partial support from an NSF grant AST-0607510.


\end{document}